\documentclass[aps,prl,twocolumn,superscriptaddress,longbibliography]{revtex4-1}
\usepackage{graphicx}
\usepackage{amsmath}
\usepackage{subfigure}
\usepackage{amsfonts}
\usepackage{amssymb}
\usepackage{mathtools}
\usepackage{siunitx}

\usepackage{color}
\definecolor{darkblue}{rgb}{0.0,0.0,0.6}

\usepackage[colorlinks = true, linkcolor = darkblue, urlcolor  = darkblue, citecolor = darkblue, anchorcolor = darkblue]{hyperref}

\newcommand{\ie}{i.\,e.\ }

\newcommand{\fref}[1]{\text{Fig.}~\ref{#1}}

\usepackage{color}
\usepackage[utf8]{inputenc}

\begin{document}

\title{Supersolid properties of a Bose-Einstein condensate in a ring resonator}
\author{S. C. Schuster} \email[]{simon.schuster@uni-tuebingen.de}
\affiliation{Physikalisches Institut, Eberhard-Karls-Universit\"{a}t T\"{u}bingen, Auf der
Morgenstelle 14, D-72076 T\"{u}bingen, Germany}
\author{P. Wolf}
\affiliation{Physikalisches Institut, Eberhard-Karls-Universit\"{a}t T\"{u}bingen, Auf der Morgenstelle 14, D-72076 T\"{u}bingen, Germany}
\author{S. Ostermann} 
\affiliation{Institut für Theoretische Physik, Universit\"{a}t Innsbruck, Technikerstaße 21a, A-6020 Innsbruck, Austria.}
\author{S. Slama}
\affiliation{Physikalisches Institut, Eberhard-Karls-Universit\"{a}t T\"{u}bingen, Auf der
Morgenstelle 14, D-72076 T\"{u}bingen, Germany}
\author{C. Zimmermann}
\affiliation{Physikalisches Institut, Eberhard-Karls-Universit\"{a}t T\"{u}bingen, Auf der
Morgenstelle 14, D-72076 T\"{u}bingen, Germany}

\begin{abstract}
We investigate the dynamics of a Bose-Einstein condensate interacting with two non-interfering and counterpropagating modes of a ring resonator. Superfluid, supersolid and dynamic phases are identified experimentally and theoretically. The supersolid phase is obtained for sufficiently equal pump strengths for the two modes. In this regime we observe the emergence of a steady state with crystalline order, which spontaneously breaks the continuous translational symmetry of the system. The supersolidity of this state is demonstrated by the conservation of global phase coherence at the superfluid to supersolid phase transition.  Above a critical pump asymmetry the system evolves into a dynamic run-away instability commonly known as collective atomic recoil lasing. We present a phase diagram and characterize the individual phases by comparing theoretical predictions with experimental observations.
\end{abstract}
\maketitle

Decades before the first experimental realization of Bose-Einstein condensation, the existence of supersolid states of matter,~\ie states that exhibit crystalline order but at the same time superfluid properties, was proposed~\cite{gross1957unified, andreev1969quantum, chester1970speculations, leggett1970can}. First experimental efforts with solid helium~\cite{kim2004probable} claimed indications of supersolidity, however, no conclusive proof of supersolid features of this system is available up to this point~\cite{balibar2010the,boninsegni2012colloquium}. The advent of Bose-Einstein condensation opened up a new way to investigate the intriguing effect of supersolidity. The fundamental requirements for supersolidity are the breaking of a gauge symmetry which results in superfluidity, and the spontaneous breaking of a continuous symmetry leading to crystalline order.

In recent years, different systems fulfilling the aforementioned requirements were studied theoretically and experimentally. This resulted in the first experimental realization of a supersolid state of matter, using a Bose-Einstein condensate (BEC) in two crossed standing wave resonators~\cite{leonard2017supersolid,leonard2017monitoring} and shortly after in a spin-orbit coupled spinor BEC~\cite{li2017a}. Very recently also the droplet state of a dipolar quantum gas was reported to show supersolid properties~\cite{chomaz2019long, bottcher2019transient,natale2019excitation, Modugno}. It was predicted lately that a transversally pumped ring cavity features a seminal tool to realize a supersolid state of a BEC~\cite{mivehvar2018driven}. Here we follow a related geometry, which is expected to show even richer physics~\cite{ostermann2015atomic}, using a BEC coupling two counterpropagating pumped ring cavity modes. The modes of a ring resonator are running plane waves, which reflects the continuous translational symmetry of the system.

\begin{figure}
\includegraphics[width=7.0cm]{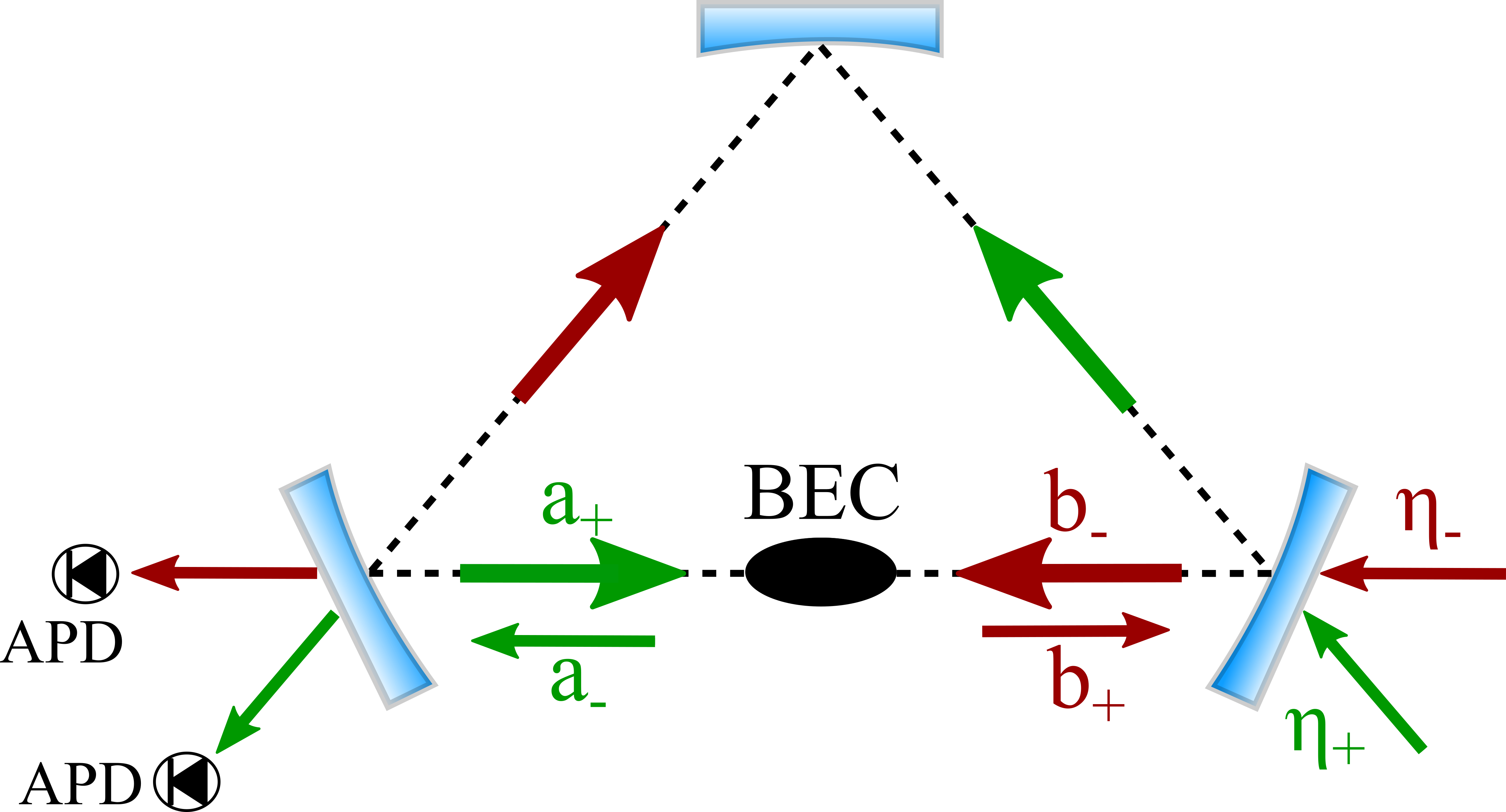}
\caption{
Experimental setup of a BEC placed in two counterpropagating modes of a ring resonator. The low finesse $TEM_{20}$ mode (green, p pol.) and the high finesse $TEM_{00}$ mode (red, s pol.) are longitudinally pumped with pump rates $\eta_\pm$. Both modes are separated in frequency by \SI{160}{\mega \Hz}. The pump power in both modes is monitored with APD's by the transmission through one of the resonator mirrors.}
\label{fig:setup}
\end{figure}

As outlined in previous work~\cite{slama2007, schmidt2014dynamical, schuster2018pinning}, a BEC in an optical ring resonator mode that is pumped longitudinally from one direction,  features an exponential instability above a critical pump power. This instability is commonly referred to as collective atomic recoil lasing (CARL)~\cite{bonifacio1994exponential, bonifacio1994collective}. In this case atomic density fluctuations scatter light from the pumped mode into its reverse mode. Interference of the pumped mode with the scattered light field creates an optical potential for the atoms and leads to atom bunching which, in turn, enhances the scattering with positive feedback. The continuous momentum transfer from scattered light onto the atoms leads to an acceleration of both the optical potential and the atomic density modulation. In this letter we show that, by pumping the atoms simultaneously from both directions with two non-interfering light fields with sufficiently equal strength, see~\fref{fig:setup}, this run-away instability is supressed. When both beams have the same pump power, a stationary atomic density modulation is realized~\cite{ostermann2015atomic}. This self-ordering phase transition breaks the ring resonator's translational symmetry leading to crystalline order while the BEC's superfluidity is maintained. Therefore, this spontaneous crystallization process is associated with the formation of a supersolid phase. In fact, this phase marks the first experimental realization of a stable phase in a ring resonator geometry. \\
In the following we experimentally and theoretically explore the first realization of a BEC coupling two non-interfering optical resonator modes. A phase diagram is recorded, distinguishing the superfluid, supersolid and dynamic (CARL) phase.  Furthermore, we discuss representative time evolution curves in the different regimes and demonstrate that global phase coherence is preserved in the emergent density modulation. Our results are based on the analysis of atomic momentum distributions, through the evaluation of time-of-flight (TOF) expansion pictures of the atoms.

The experimental setup in~\fref{fig:setup} is very similar as in~\cite{schuster2018pinning}, with the only difference being that here we use two different transversal modes to illuminate a BEC of $^{87}Rb$ from two directions. The atoms are magnetically trapped and positioned in the center intensity maxima of the pump modes. Radial and longitudinal frequencies of the magnetic trap are $(f_r,f_t)=(\SI{36.6}{\hertz},\SI{100}{\hertz})$. The pump mode in $+$-direction is a $TEM_{20}$ (Transverse Electromagnetic Mode) in $s$-polarization with a decay rate of the electric field amplitude of $\kappa_+=2\pi\cdot$\SI{18}{\kilo\hertz}. The pump mode used in $-$-direction is the $p$-polarized $TEM_{00}$ with a field decay rate of $\kappa_-=2\pi\cdot$\SI{5}{\kilo\hertz}. Additionally they are separated in frequency by $\SI{160}{\mega\hertz}$, thus any possible interference and polarization grating of both modes averages out on the millisecond-time scale of the experiment. The decay rate $\kappa_+$ is slightly bigger then the two photon recoil frequency $\omega_r=2\pi\cdot$\SI{14.51}{\kilo\hertz}, but no qualitatively different dynamics is expected compared to $\kappa_-$. The frequency of the pump light fields is detuned $\Delta_a=\omega_\pm-\omega_0=-2\pi\cdot$\SI{63}{\giga\hertz} relative to the atomic transition frequency $\omega_0$ ($D1$ Line: $5s_{1/2}$, $F=2$ to $5p_{1/2}$, $F=2$). Both pump frequencies $\omega_\pm$ are set to a fixed detuning by $\Delta_{c}=\omega_{c\pm}-\omega_\pm=1\omega_r$, with the respective resonance frequencies of the $TEM_{20}$ and $TEM_{00}$ modes $\omega_{c\pm}$. The detunings $\Delta_c$ and $\Delta_a$ were chosen such, in order to stay well below the Pinning-Transition observed in a previus work \cite{schuster2018pinning}. The absorption images of the BEC are taken after \SI{25}{\milli\second} of ballistic flight to determine the occupations $|c_n|^2$ of the individual BEC momentum states, which are normalized to $\sum_n |c_n|^2=1$.
\begin{figure}
\includegraphics[width=0.5\textwidth]{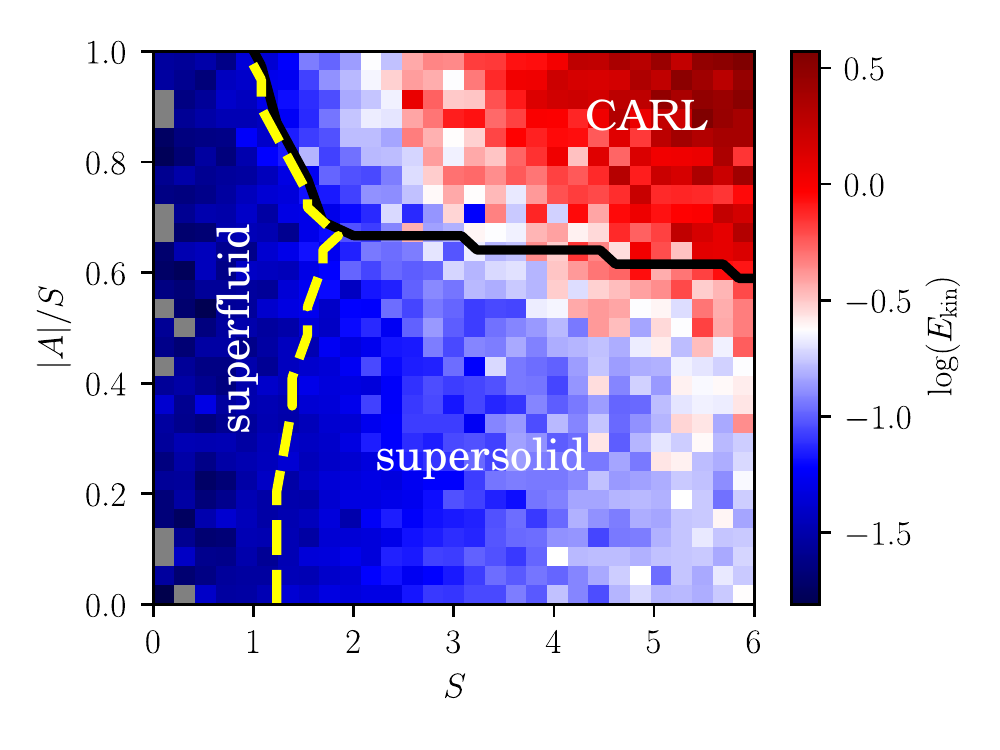}
\caption{Phase diagram presented as logarithmic plot of the total kinetic energy of the BEC after an interaction time of $\SI{1}{\milli\second}$. By numerical analysis of \eqref{eqn:theo_model}, we are able to distinguish three fundamentally different phases: the superfluid phase where the BEC has a homogeneous density distribution (to the left of the dashed yellow line), a supersolid phase (to the right of the dashed yellow and below the solid black line) and a CARL regime (above the solid black line). The grey squares mark parameter sets where no measurements were performed.}
\label{fig:phase_diagram}
\end{figure}
In~\fref{fig:phase_diagram} the observed phase diagram for different pump powers $S$ and pump asymmetries $|A|/S$ is shown. The pump parameters are defined as
\begin{equation}
S=\frac{|a_+|^2}{|a_+^{\textrm{crit}}|^2}+\frac{|b_-|^2}{|b_-^{\textrm{crit}}|^2}
\end{equation}
and
\begin{equation}
A=\frac{|a_+|^2}{|a_+^{\textrm{crit}}|^2}-\frac{|b_-|^2}{|b_-^{\textrm{crit}}|^2}.
\end{equation}
$|a_+^{\textrm{crit}}|^2$ and $|b_-^{\textrm{crit}}|^2$ are the respective threshold pump powers for which, at full asymmetric pumping (counter propagating pump power equals zero), the condensate population is depleted by $10\%$ due to the CARL process. To distinguish the different phases, we calculate the atomic kinetic energy $E_\textrm{kin}=\sum_n n^2 |c_n|^2$ after an interaction time of $\SI{1}{\milli\second}$.  Each square in ~\fref{fig:phase_diagram} represents the average of up to 20 data points, with a total of 6400 single measurements and an average atom number of $N=$\SI{3.6e5}{}$\pm$\SI{6.8e4}{}.\\
We identify three fundamentally different phase regions. For pump strengths below the critical intensity (left of the yellow dashed line in~\fref{fig:phase_diagram}) no modulation of the BEC is observed and we call this the superfluid phase. For larger pump asymmetries $|A|/S$ and sufficiently strong pumping the BEC enters the CARL regime (above solid black line), in which the center of mass is accelerated, thus higher momentum states are occupied and the highest kinetic energy is reached. In contrast, for smaller pump asymmetries the system enters the supersolid phase (region below black solid line and to the right of the yellow dashed line in~\fref{fig:phase_diagram}). In this phase, although both counterpropagating pump modes do not interfere and thus cannot generate an optical lattice in which the BEC density could be structured, a stable atomic density modulation emerges, breaking the  system's continuous $U(1)$ symmetry.
\begin{figure*}
\includegraphics[width=0.9\textwidth]{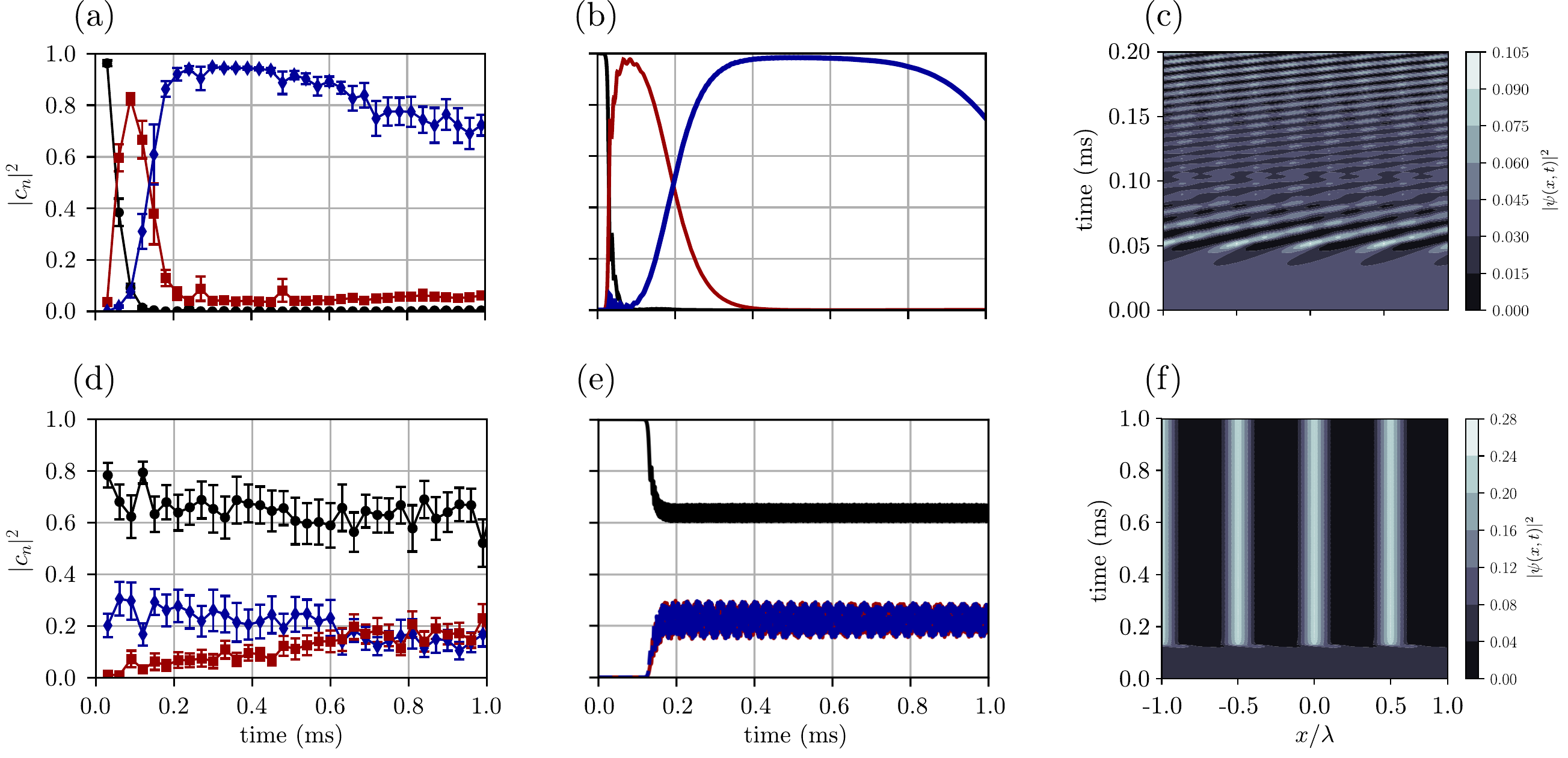}
\caption{Time evolution in the CARL regime (a-c) for $(|A|/S,S)=(0.95,6.32)$ and in the supersolid regime (d-f) for $(|A|/S,S)=(0.008,8.15)$. 
The first column shows the measured evolution of the momentum states $|c_n|^2$  with $n=0, 1, 2$ (black, red, blue) in the CARL regime (a) and $n=-1, 0, 1$ (blue, black, red) in the supersolid regime (d). The errorbars are the standard deviation of the mean value of each datapoint. The continuous lines in the second column show numerical simulation of the equation of motion~\eqref{eqn:theo_model} with the same parameters as in the corresponding measurement in the first column. For roughly symmetric pump powers a stationary momentum distribution is observed after $\SI{600}{\micro\second}$. In the third column, the numerically obtained time evolution of the spatial distribution of the BEC for the two different regimes CARL (c) and quasi stationary supersolid (f) are presented.}\label{fig:evolution}
\end{figure*}
This crystallization process  is directly related to a gapless Goldstone mode which appears above this critical pump strength; see Supplemental Material for details. For symmetric pumping ($|A|/S=0$) the density modulation is expected to be stationary, corresponding to a symmetric occupation of momentum states around $n=0$. Asymmetric pumping in the supersolid phase results in a density modulation moving with \emph{constant} velocity, associated with an asymmetric momentum distribution. For a given pump asymmetry, the system realizes a stable momentum distribution in which the net force onto the BEC's center of mass disappears. For a certain detuning with $\Delta_c>0$, the probability of a pump photon being scattered from an atom that is moving towards it, is greater then being scattered on an atom that is moving away from the pump beam. This mechanism, similar to cavity cooling~\cite{gangl2000cold,Wolke2012,Sandner2013}, is stabilizing the density modulation, having the atoms scattered back into the center of the momentum distribution, and is compensating the asymmetric pump powers.\\ 
The dynamics of the system is well described by a one dimensional mean-field model where $x$ denotes the cavity axis. The BEC dynamics is governed by the time-dependent equation (note that we neglect local particle-particle interactions)
\begin{subequations}
\begin{align}
i\partial_t\psi(x,t)=H_\mathrm{mf}\psi(x,t)
\label{eqn:GPE}
\end{align}
with the mean-field Hamiltonian
\begin{multline}
H_\mathrm{mf}=-\frac{1}{2m}\frac{\partial^2}{\partial x^2} +\hbar U_0 \Big(e^{-2ikx}a^*_+a_-+e^{2ikx}a_-^*a_+\\
+e^{-2ikx}b_+^*b_-+e^{2ikx}b_-^*b_+\Big),
\end{multline}
with the cavity wavevector $k=2\pi/\lambda$ ($\lambda$ is the cavity resonance wavelength). The mode dynamics is governed by the set of equations
\begin{align}
\partial_ta_+&=-\left(i\delta_c+\kappa_+\right)a_+-iU_0\mathcal{N}a_-+\eta_+\\
\partial_ta_-&=-\left(i\delta_c+\kappa_+\right)a_--iU_0\mathcal{N}^*a_+\\
\partial_tb_+&=-\left(i\delta_c+\kappa_-\right)b_+-iU_0\mathcal{N}b_-\\
\partial_tb_-&=-\left(i\delta_c+\kappa_-\right)b_--iU_0\mathcal{N}^*b_++\eta_-,
\end{align}
\label{eqn:theo_model}
\end{subequations}
where we introduced $\delta_c\coloneqq\Delta_c+NU_0$ and the bunching parameter $\mathcal{N}\coloneqq\int dx e^{-2ikx}|\psi(x,t)|^2$. The particle number is denoted as $N$, $\hbar U_0$ is the potential depth per photon and $\eta_\pm$ are the effective pump amplitudes. Note that there is no direct interaction between the modes $a_\pm$ and $b_\pm$ because of their orthogonal polarizations. They only interact indirectly via the BEC density. In addition, the set of equations~\eqref{eqn:theo_model} reflects the continuous translational symmetry of the considered system,~\ie they are invariant under spatial translations $x\rightarrow x+\Delta x$ since these can be compensated by phase shifts $a_\pm\rightarrow a_\pm e^{\mp i 2 k\Delta x}$. The set of equations~\eqref{eqn:theo_model} allows us to estimate a numerical phase boundary between the supersolid and CARL regime (dashed yellow and solid black curves in~\fref{fig:phase_diagram}) by simulating the experimental procedure and expanding the BEC wavefunction into momentum eigenstates via $\psi(x,t)=\sum_n c_n(t) e^{2 i n k x}$. For a fixed $|A|/S$, the dashed yellow line marks the critial $S$ at which the resting BEC ($n=0$ momentum state) is depleted by $10\%$. Likewise, for a fixed $S$, the solid black line marks the critical asymmetry parameter $|A|/S$ at which the resting condensate is depleted by $10\%$. We find good agreement between the measured data and the numerical estimations.

Let us now focus on the analysis of some representative time evolution measurements in the two non-trivial phases of the phase diagram (see~\fref{fig:evolution}). In the CARL regime [\fref{fig:evolution}(a-c)] the atoms are quickly transferred to higher momenta due to the exponential instability. This means that after a time evolution of around \SI{5}{\micro\second}, the density modulation forms up and accelerates. In contrast, in the supersolid regime [\fref{fig:evolution}(d-f)], a stable and close to symmetrical occupation is observed after around \SI{600}{\micro\second}. The slower observed settling time in~\fref{fig:evolution}(d) can be explained by heating introduced by the pump fields during the switch-on process, which is not contained in the theoretical model. The numerical curves obtained from~\eqref{eqn:theo_model} show good agreement with the experimental data. Figs.~\ref{fig:evolution}(c) and (f) show the time evolution of the spatial density distribution $|\psi(x,t)|^2$ calculated from the corresponding numerical time curves in Figs.~\ref{fig:evolution}(b) and (e). They nicely illustrate the difference between stable crystalline order (f) and the accelerating density modulation in the CARL regime (c).   
\begin{figure}
\includegraphics[width=0.4\textwidth]{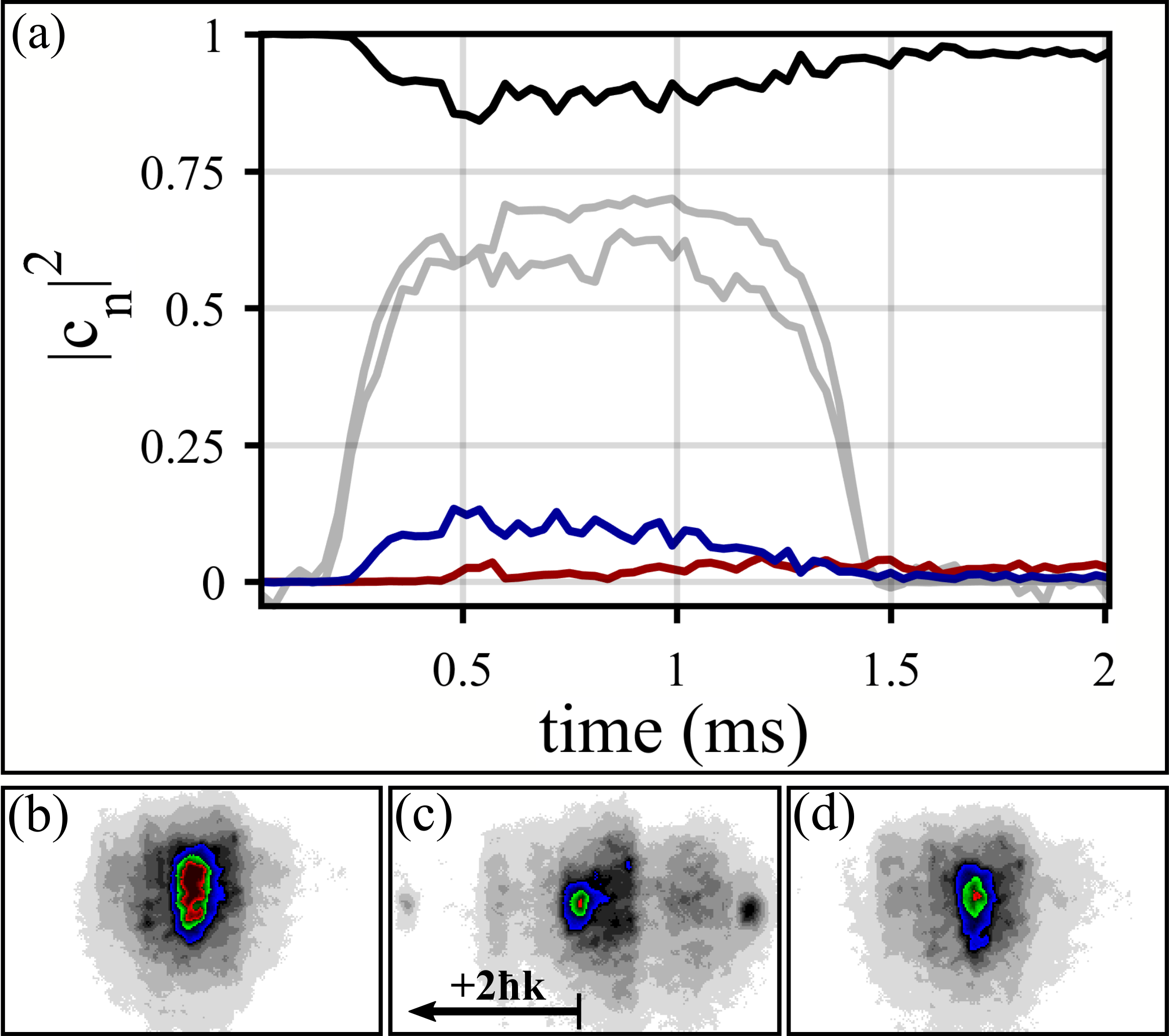}
\caption{Reversibility and phase coherence of the excited crystalline structure is demonstrated, by symmetrically ramping the pump power for $(|A|/S,S)=(0.06,8.2)$. In (a) the evolution of single momentum states $n=-1, 0, 1$ (blue, black, red) is presented together with the exemplary progression of both pump powers (grey), while ramps are applied. (b-d) show the average of 8 TOF pictures for pump times (b) $t<\SI{0.2}{\milli\second}$, (c) $\SI{0.5}{\milli\second}<t< \SI{1}{\milli\second}$ and (d) $t>\SI{1.5}{\milli\second}$. The TOF pictures are scaled alike.} 
\label{fig:ramps}
\end{figure}

Up to now we have only quantified the stable density modulation in the supersolid state, which is a direct result of breaking the continuous symmetry. An important condition for supersolidity is however the conservation of global phase coherence during the phase transition from the homogeneous phase to peridic order \cite{chomaz2019long}. In order to analyse the phase coherence of the supersolid state we perform ramping mesurements, similar to those proposed in \cite{Ostermann2017}. We start in the superfluid phase with symmetric pumping and increase the pumping strengths in order to reach the supersolid regime. After a short waiting time the pumping strengths are reduced to the inital value. An exemplary time evolution is shown in~\fref{fig:ramps}. 
The crystallization process adiabatically follows the ramp applied during the first \SI{500}{\micro \second}, synchronously occupying the $n=\pm 1$ momentum states at $\pm 2\hbar k$ (see~\fref{fig:ramps}(a) and (c)). 
After the inverse ramp, starting at \SI{1}{\milli \second}, we find the BEC almost completely restored [\fref{fig:ramps}(d)] which is a clear indication that global phase coherence was maintained during the transitions between the superfluid and supersolid phase.  
Rayleigh scattering and a non ideally adiabatic ramp may lead to heating and being responsible for the incomplete reversal.     

In conclusion, we have investigated an atomic BEC in an optical ring resonator which couples two non-interfering counterpropagating pump fields. The phase diagram exhibits three fundamentally different regions as predicted in previous work~\cite{ostermann2015atomic}. In particular, for sufficiently symmetric pump configurations the system undergoes a phase transition to a supersolid state of matter as it was proposed recently~\cite{mivehvar2018driven}. The supersolidity in this particular system solely relies on the relative phase between the modes and the global phase of the condensate. This implies that the realized supersolid state is very robust against dissipation (particle loss or photon loss), which marks a big advantage compared to other geometries realizing supersolidity. In addition, the periodic ordering results from the long-range interactions induced by the coherent resonator field. This suppresses decoherent processes and, thus, results in the good conservation of global phase coherence of the BEC at the superfluid to supersolid phase transition.

Apart from the possibility to study the properties of the supersolid state, the realized geometry constitutes a seminal tool for a variety of future applications. Recently the supersolid state in a ring resonator was predicted to have applications in high-precission metrology~\cite{gietka2019supersolid}.  In addition, the stabilization mechanism which leads to the the stable supersolid phase could be a very efficient cooling mechanism for atom clouds~\cite{gangl2000cold}. The studied geometry can also be generalized to include spin degrees of freedom by using a spinor BEC in a similar geometry~\cite{ostermann2019cavity}. In this case topological phase transitions and spin waves with supersolid properties due to cavity-induced spin-orbit coupling are expected to emerge.\\

We acknowledge helpful discussion with Helmut Ritsch. This work has been supported by the Deutsche
Forschungsgemeinschaft  (ZI 419/8-1) and by the FWF project I 3964-N27.

\bibliography{arxiv_submission_supersolid}

\newpage
\widetext
\setcounter{equation}{0}
\setcounter{figure}{0}
\renewcommand{\theequation}{S\arabic{equation}}
\renewcommand{\thefigure}{S\arabic{figure}}

\section{Supplemental Material}
Here we perform some additional analysis of the system. In particular we focus on the calculation of the collective excitation spectrum resulting in the gapless Goldstone mode and give an analytical threshold pump strength for the formation of a supersolid.

\section{Collective Excitations - Goldstone mode}
\begin{figure}[b]
\centering
\includegraphics[width=0.6\textwidth]{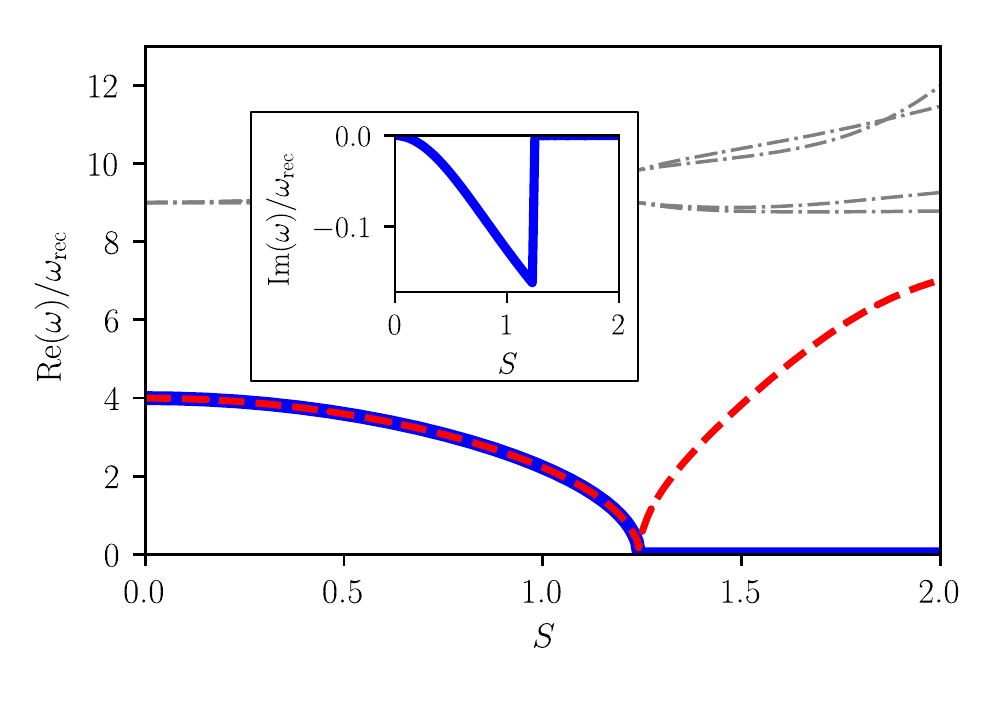} 
\caption{Collective excitation spectrum. The lowest lying excitation branch (blue curve) shows the emergence of a gapless Goldstone mode due to the spontaneous breaking of the contiunuous $U(1)$ symmetry at the critical pump strength. The inset shows the imaginary part of the lowest excitation branch, which is zero in the supersolid regime. This indicates the vanishing of friction forces along the resonator axis on the supersolid state.}
\label{fig:coll_ex}
\end{figure}

To find the collective excitation spectrum for the supersolid we linearize the set of equations given in the main text~\eqref{eqn:theo_model} around the stationary mean-field solutions $\psi_0(x)$. We perform the ansatz $\psi(x,t)=\psi_0(x)+\delta\psi(x,t)$, $\alpha_\pm(t)=\alpha_0^\pm+\delta\alpha_\pm(t)$ and $\beta_\pm(t)=\beta_0^\pm+\delta\beta_\pm(t)$ and keep only linear terms in $\delta\psi$, $\delta\alpha_\pm$ and $\delta\beta_\pm$. The linearized equations for the BEC dynamics after a Bogoliubov transformation $\delta\psi(x,t)=\delta\psi^{(+)}(x)e^{-i\omega t}+[\delta\psi^{(-)}(x)]^*e^{i\omega^* t}$ read
\begin{subequations}
\begin{align}
\omega \delta\psi^{(+)}=\frac{1}{\hbar}\left(H_\mathrm{mf}-\mu\right)\delta\psi^{(+)}&+U_0\left(A_{+*}^*\delta\alpha_+^{(+)}+A_{-*}^*\delta\alpha_-^{(+)}+A_+\delta\alpha_+^{(-)}+A_-\delta\alpha_-^{(-)}\right)\\\nonumber
&+U_0\left(B_{+*}^*\delta\beta_+^{(+)}+B_{-*}^*\delta\beta_-^{(+)}+B_+\delta\beta_+^{(-)}+B_-\delta\beta_-^{(-)}\right)\\
\omega \delta\psi^{(-)}=-\frac{1}{\hbar}\left(H_\mathrm{mf}-\mu\right)^*\delta\psi^{(-)}&-U_0\left(A_{+*}\delta\alpha_+^{(-)}+A_{-*}\delta\alpha_-^{(-)}+A_+^*\delta\alpha_+^{(+)}+A_-^*\delta\alpha_-^{(+)}\right)\\\nonumber
&-U_0\left(B_{+*}\delta\beta_+^{(-)}+B_{-*}\delta\beta_-^{(-)}+B_+^*\delta\beta_+^{(+)}+B_-^*\delta\beta_-^{(+)}\right),
\end{align}
\label{eqn:lin_atoms}
\end{subequations}
\noindent with $A_\pm:=\psi_0(\alpha_0^\pm+e^{\mp2ikx}\alpha_0^\mp)$, $A_{\pm,*}:=\psi_0^*(\alpha_0^\pm+e^{\mp2ikx}\alpha_0^\mp)$, $B_\pm:=\psi_0(\beta_0^\pm+e^{\mp2ikx}\beta_0^\mp)$ and $B_{\pm,*}:=\psi_0^*(\beta_0^\pm+e^{\mp2ikx}\beta_0^\mp)$.

The linearized equations for the cavity modes $\alpha_\pm$ after the transformation $\alpha_\pm(t)=\delta\alpha_\pm^{(+)}e^{-i\omega t}+[\delta\alpha_\pm^{(-)}]^*e^{i\omega^*t}$ are given as
\begin{subequations}
\begin{align}
\omega\delta\alpha_+^{(+)}&=-\Delta_\mathrm{LF}\delta\alpha_+^{(+)}+U_0\mathcal{N}\delta\alpha_-^{(+)}+U_0\mathcal{A}_{+*}\delta\psi^{(+)}+U_0\mathcal{A}_+\delta\psi^{(-)}\\
\omega\delta\alpha_+^{(-)}&=\Delta_\mathrm{LF}^*\delta\alpha_+^{(-)}-U_0\mathcal{N}^*\delta\alpha_-^{(-)}-U_0\mathcal{A}_{+*}^*\delta\psi^{(-)}-U_0\mathcal{A}_+^*\delta\psi^{(+)}\\
\omega\delta\alpha_-^{(+)}&=-\Delta_\mathrm{LF}\delta\alpha_-^{(+)}+U_0\mathcal{N}^*\delta\alpha_+^{(+)}+U_0\mathcal{A}_{-*}\delta\psi^{(+)}+U_0\mathcal{A}_-\delta\psi^{(-)}\\
\omega\delta\alpha_-^{(-)}&=\Delta_\mathrm{LF}^*\delta\alpha_-^{(-)}-U_0\mathcal{N}\delta\alpha_+^{(-)}-U_0\mathcal{A}_{-*}^*\delta\psi^{(-)}-U_0\mathcal{A}_-^*\delta\psi^{(+)},
\end{align}
\label{eqn:lin_mode_a}
\end{subequations}
where we introduced $\mathcal{A}_\pm\xi:=\int dx A_\pm \xi$, $\mathcal{A}_{\pm*}\xi:=\int dx A_{\pm*} \xi$ and $\Delta_\mathrm{LF}=-\Delta_c+i\kappa_+-NU0$.

For the cavity modes $\beta_\pm$ again performing the transformation $\delta\beta_\pm(t)=\delta\beta_\pm^{(+)}e^{-i\omega t}+[\delta\beta_\pm^{(-)}]^*e^{i\omega^* t}$ leads to
\begin{subequations}
\begin{align}
\omega\delta\beta_+^{(+)}&=-\Delta_\mathrm{HF}\delta\beta_+^{(+)}+U_0\mathcal{N}\delta\beta_-^{(+)}+U_0\mathcal{B}_{+*}\delta\psi^{(+)}+U_0\mathcal{B}_+\delta\psi^{(-)}\\
\omega\delta\beta_+^{(-)}&=\Delta_\mathrm{HF}^*\delta\beta_+^{(-)}-U_0\mathcal{N}^*\delta\beta_-^{(-)}-U_0\mathcal{B}_{+*}^*\delta\psi^{(-)}-U_0\mathcal{B}_+^*\delta\psi^{(+)}\\
\omega\delta\beta_-^{(+)}&=-\Delta_\mathrm{HF}\delta\beta_-^{(+)}+U_0\mathcal{N}^*\delta\beta_+^{(+)}+U_0\mathcal{B}_{-*}\delta\psi^{(+)}+U_0\mathcal{B}_-\delta\psi^{(-)}\\
\omega\delta\beta_-^{(-)}&=\Delta_\mathrm{HF}^*\delta\beta_-^{(-)}-U_0\mathcal{N}\delta\beta_+^{(-)}-U_0\mathcal{B}_{-*}^*\delta\psi^{(-)}-U_0\mathcal{B}_-^*\delta\psi^{(+)},
\end{align}
\label{eqn:lin_mode_b}
\end{subequations}
\noindent with $\mathcal{B}_\pm\xi:=\int dx B_\pm \xi$, $\mathcal{B}_{\pm*}\xi:=\int dx B_{\pm*} \xi$ and $\Delta_\mathrm{HF}:=-\Delta_c+i\kappa_--NU0$.

This set of equations can be written in matrix form $\omega\mathbf{f}=\mathbf{M}_B\mathbf{f}$ with $\mathbf{f}=(\delta\psi^{(\pm)},\delta\alpha_\pm^{(\pm)},\delta\beta_\pm^{(\pm)})^T$ where the Bogoliubov matrix reads
\begin{equation}
\mathbf{M}_B=
\begin{pmatrix}
\frac{1}{\hbar}(H_\mathrm{mf}-\mu) & 0 & U_0 A_{+*}^* & U_0 A_+ & U_0 A_{-*}^* & U_0 A_- & U_0 B_{+*}^* & U_0 B_+ & U_0 B_{-*}^* & U_0 B_-\\
0 & -\frac{1}{\hbar}(H_\mathrm{mf}-\mu)^* & -U_0 A_{+}^* & -U_0 A_{+*} & -U_0 A_{-}^* & -U_0 A_{-*} & -U_0 B_{+}^* & -U_0 B_{+*} & -U_0 B_{-}^* & -U_0 B_{-*}\\
U_0\mathcal{A}_{+*}  & U_0\mathcal{A}_{+} & -\Delta_\mathrm{LF} & 0 & U_0 \mathcal{N} & 0 & 0 & 0 & 0 & 0 \\
-U_0\mathcal{A}_{+}^*  & -U_0\mathcal{A}_{+*}^* & 0 & \Delta_\mathrm{LF}^* & 0 & -U_0 \mathcal{N}^* & 0 & 0 & 0 & 0 \\
U_0\mathcal{A}_{-*}  & U_0\mathcal{A}_{-} & U_0 \mathcal{N}^* & 0 & -\Delta_\mathrm{LF} &  0 & 0 & 0 & 0 & 0\\
-U_0\mathcal{A}_{-}^*  & -U_0\mathcal{A}_{-*}^* & 0 & -U_0 \mathcal{N} & 0 & \Delta_\mathrm{LF}^* & 0 & 0 & 0 & 0\\
U_0\mathcal{B}_{+*}  & U_0\mathcal{B}_{+} & 0 & 0 & 0 & 0 & -\Delta_\mathrm{HF} & 0 & U_0\mathcal{N} & 0\\
-U_0\mathcal{B}_{+}^*  & -U_0\mathcal{B}_{+*}^* & 0 & 0 & 0 & 0 & 0 & \Delta_\mathrm{HF}^* &  0 & -U_0\mathcal{N}^*\\
U_0\mathcal{B}_{-*}  & U_0\mathcal{B}_{-} & 0 & 0 & 0 & 0 & U_0\mathcal{N}^* & 0 & -\Delta_\mathrm{HF} & 0\\
-U_0\mathcal{B}_{-}^*  & -U_0\mathcal{B}_{-*}^* & 0 & 0 & 0 & 0 & 0 & -U_0\mathcal{N} &  0 & \Delta_\mathrm{HF}^*
\end{pmatrix}.
\label{eqn:MB}
\end{equation}

To find the collective excitation spectrum we calculate the selfconsistent groundstate for different values of $\eta_\pm$ numerically. For the sake of simplicity we assume equal decay rates for the two counterpropagating modes, \ie $\kappa_+=\kappa_-\equiv\kappa$ and $\Delta_\mathrm{LF}=\Delta_\mathrm{HF}\equiv\Delta$. In the experiment the two modes have different decay rates, but this assumption does not affect the fundamental physics which is discussed here. The self-consistent method only works if the pump asymmetry $A=0$ which under the performed assumptions implies $\eta_+=\eta_-$.

Diagonalizing the Bogoliubov Matrix~\eqref{eqn:MB} after numerically finding the groundstate $\psi_0(x)$ leads to the collective excitation spectrum shown in~\fref{fig:coll_ex}. We see that the lowest two modes soften at the phase transition point. One of the two modes (blue curve in~\fref{fig:coll_ex}) remains zero in the selfordered phase, which corresponds to the gapless Goldstone mode of the system. It is related to the breaking of the continuous translational symmetry. Another fundamental feature of the system is that the supersolid phase is robust against dissipation. This can be seen from the fact that the imaginary part of the Goldstone mode is zero in the supersolid regime (see inset in~\fref{fig:coll_ex}). Due to the cavity decay rate $\kappa$ the collective excitation energies can acquire an imaginary part which would correspond to damping of the corresponding excitation branch. The vanishing of the imaginary part of the Goldstone mode in the supersolid remine means that this lowest lying excitation is not damped. Therefore, the center of mass of the supersolid can move without friction along the cavity axis.

\section{Analytical Threshold}
The collective excitations calculated in the previous chapter allow us to calculate an analytical threshold for the case $A=0$. Therefore, we analyze the stability of the homogeneous BEC $\psi_0=\sqrt{N/\lambda}$ which implies $\alpha_0^-=\beta_0^+=0$, $\mu=0$ and $\mathcal{N}=0$. We perform an ansatz for the atomic perturbation $\delta\psi^{(\pm)}$ assuming that the fundamental coupling processes only take into account $0\hbar k $ and $\pm 2 \hbar k$ momenta, \ie $\delta \psi^{(\pm)}=\delta \psi_0^{(\pm)}+\delta\psi_+^{\pm}e^{2ikx}+\delta\psi_-^{(\pm)}e^{-2ikx}$.  For the Bogoliubov matrix on this subspace $\tilde{M}_B$ the characteristic equation is given via $\mathrm{Det}(\tilde{M}_B-\omega \mathbf{1})=0$ results in
\begin{equation}
8 N U_0^2(|\alpha_0^+|^2+|\beta_0^-|^2)\omega_\mathrm{rec}+(\Delta + \omega) (\omega^2-16\omega_\mathrm{rec})=0.
\label{eqn:char_eq}
\end{equation}
The real part of the zeroth order solution ($\omega=0$) gives the critical cavity field amplitudes
\begin{equation}
|\alpha_0^+|_c^2+|\beta_0^-|_c=\frac{2\omega_\mathrm{rec}|-\Delta_c-N U_0|}{NU_0^2}.
\label{eqn:alpha_thres}
\end{equation}
The absolute values for the steady state solutions are given as
\begin{equation}
|\alpha_{\mathrm{ss}}^+|=\frac{|\eta_+|^2}{(-\Delta_c-NU_0)^2+\kappa^2}, \quad \quad |\beta_{\mathrm{ss}}^+|=\frac{|\eta_-|^2}{(-\Delta_c-NU_0)^2+\kappa^2}.
\end{equation}
Plugging this into~\eqref{eqn:alpha_thres} leads to the critical condition for the pump strengths
\begin{equation}
\tilde{S}_c:=|\eta_{c, +}|^2+|\eta_{c, -}|^2=\frac{((-\Delta_c-N U_0)^2+\kappa^2)^2}{NU_0^2|-\Delta_c-NU_0|}2\omega_\mathrm{rec}.
\label{eqn:eta_thres}
\end{equation}

\end{document}